\documentclass[12pt,preprint]{aastex}

\begin{document}

\title{ A comparative study of the X-ray afterglow properties of optically bright
and dark GRBs }

\author{ M. De Pasquale \altaffilmark{1,2}, L. Piro \altaffilmark{1}, R. Perna\altaffilmark{1,3},
E. Costa\altaffilmark{2}, M. Feroci\altaffilmark{2}, G.
Gandolfi\altaffilmark{2}, J. in 't Zand\altaffilmark{4}, L.
Nicastro\altaffilmark{5}, F. Frontera\altaffilmark{6,7}, L. A.
Antonelli\altaffilmark{8}, F. Fiore \altaffilmark{8}, G.
Stratta\altaffilmark{2,8}. }

\altaffiltext{1}{ IASF - CNR, Via Fosso Del Cavaliere 100,
I-00133, Rome, Italy } \altaffiltext{2}{University "La Sapienza",
Piazzale Aldo Moro 5, I-00185 Rome, Italy} \altaffiltext{3}{
Harvard-Smithsonian Center for Astrophysics, 60 Cambridge Street,
MA02138 USA } \altaffiltext{4}{ Space Research Organization
Netherlands, Sorbonnelaan 2, 3584 CA Utrecht, The Netherlands }
\altaffiltext{5}{ IASF - CNR, Via Ugo la Malfa 153, I-90146
Palermo, Italy } \altaffiltext{6}{ IASF - CNR, Via Gobetti 101,
I-40129 Bologna, Italy} \altaffiltext{7}{ Physics Department,
University of Ferrara, Via Paradiso 11, I-44100, Italy}
\altaffiltext{8} {Rome Astronomical Observatory, Via di Frascati
33, I-00044 Italy }

\begin{abstract}

 We have examined the complete set of X-ray afterglow observations
of dark and optically bright GRBs performed by BeppoSAX until
February 2001.
 X-ray afterglows are detected in $\sim 90\%$ of the cases.
 We do not find significant differences in the X-ray spectral
shape, in particular no higher X-ray absorption in GRBs without
optical transient ( dark GRBs ) compared to GRBs with optical
transient ( OTGRBs ). Rather, we find that the 1.6-10 keV flux of
OTGRBs is on average about 5 times larger than that of the dark
GRBs. A K-S test shows that this difference is significant at
99.8\% probability. Under the assumption that dark and OTGRB have
similar spectra, this could suggest that the first are uncaught in
the optical band because they are just faint sources. In order to
test this hypothesis, we have determined the optical-to-X ray flux
ratios of the sample. OTGRBs show a remarkably narrow distribution
of flux ratios, which corresponds to an average optical-to-x
spectral index $\overline{\alpha}_{OX} ^{ OT}= 0.794\pm 0.054 $.
We find that, while 75\% of dark GRBs have flux ratio upper limits
still consistent with those of OT GRBs, the remaining 25\% are 4 -
10 times weaker in optical than in X-rays. The significance of
this result is $\geq 2.6\sigma$. If this sub-population of dark
GRBs were constituted by objects assimilable to OTGRBs, they
should have shown optical fluxes higher than upper limits actually
found. We discuss the possible causes of their behaviour,
including a possible occurrence in high density clouds or origin
at very high redshift and a connection with ancient, Population
III stars.

\end{abstract}

\section{Introduction}

 About 50\% of well-localized GRBs show optical transients (OTs)
successive to the prompt gamma-ray emission, whereas an X-ray counterpart is present in 90\% of cases.  It is
possible that late and shallow observations could not detect the OTs in some cases; several authors argue that
dim and/or rapid decaying transients could bias the determination of the fraction of truly obscure GRBs
\citep{Fyn01a,Ber02}. However, recent reanalysis of optical observations \citep{Rei01,Ghi00,Laz00} has shown
that GRBs without OT detection (usually dark GRBs, FOAs Failed Optical Afterglows, or GHOSTs, Gamma ray burst
Hiding an Optical Source Transient) have had on average weaker optical counterparts, at least 2 magnitudes in
the R band, than GRBs with OTs. Therefore, they appear to constitute a different class of objects, albeit there
could be a fraction undetected for bad imaging.

 Two hypothesis have been put forward to explain the behaviour of GHOSTs.
First, they are similar to the other bright GRBs, except for the
fact that their lines of sight pass through large and dusty
molecular clouds, that cause high absorption. Second, they are
more distant than GRBs with OT, at $ z \ga 5 $ \citep{Fru99}, so
that the Lyman break is redshifted into the optical band. These
GRBs might be associated with the explosion of ancient Population
III, high mass stars.
 Nevertheless, the distances of a few dark GRBs have been determined
and they do not imply high redshifts \citep{Djo02,Ant00,Pir02}

 Goal of this paper is an analysis of a complete sample of
BeppoSAX X-ray afterglows in order to distinguish between these
various scenarios, including all x-ray fast observation from the
launch to February 2001. In \S 2 and \S 3 we present the data
analysis of the afterglows and we show the results, whose
implications are discussed in \S 4. Finally, we summarize our
conclusions in \S 5.

\section{ Data Analysis }

 We have analyzed all the 31 fast BeppoSAX observations of GRB X-ray afterglows taken by the
Low Energy (0.1 - 10 keV) and Medium Energy (1.6 - 10 keV)
Concentrator Spectrometer (LECS and MECS respectively, see Parmar
et al. 1997, Boella et al. 1997) up to GRB010222. We excluded only
GRB960720 for the late follow up, GRB990705 due to its high
contamination of a nearby X-ray source, and GRB980425 due to its
peculiarity. X-ray follow up observations usually start $\sim 9 -
10$ hours after the high energy event and the typical observation
time is $\sim 2\times 10^{5}$ seconds for MECS and $\sim 5 \times
10^{4} $ for LECS. The exposure - or integration - lasts $\sim
1/3$ of the observation.

 In order to find out the GRB X-ray afterglows, we first have built
up the images of each GRB with the MECS and selected sources with
$3\sigma$ significance within the WFC error box. Successively, we
have built the light curves of these sources to recognize
afterglows through their typical fading emission. The counts have
been collected within a circle centered on source with radius $ r
= 4 $ arcmin. Then we have subtracted the background collected in
annuli around the extraction area and 5 times more extended. Local
background have been used in order to take into account possible
time fluctuations. TOOs successive to the first one (typically
$\sim 2$ day after) have been used, if available.
 We have fitted the light curves with simple power law
$N_{cts} \varpropto t^{- \delta}$ (where $N_{cts}$ is counts per
second) and 26 sources with decaying index $\delta >0$ (at $90\%$
Confidence Level) have been recognized as GRB afterglows.
 In the case of GRB970111, GRB991106 and GRB000615 we have detected 1 source within WFC
error box that does not show a significant fading behaviour. We will refer to them as "candidate"
afterglows\footnote{In the case of GRB991106, the source in the WFC error box could be a type-I Galactic X-ray
burst \citep{Cor02}.}. We have calculated the probability to have serendipitous sources with their flux within
the WFC error box, adopting the Log N - Log S distribution for BeppoSAX released by \citet{Gio96}. The
probability is $\cong 0.027$ for each one, while the probability that all of them are not afterglows is $P\sim
10^{-5}$.

 The MECS integration time for GRB990907 was only 1070 seconds, so that the presence of a fading flux
could not be verified. The X-ray source detected was recognized as the GRB afterglow because
the probability to have a serendipitous source with flux $ 10^{-12}$ erg cm$^{-2}$ sec$^{-1}$
(see further) in the WFC error box was $ \simeq 0.007 $.

 Finally, in the case of GRB990217 and GRB010220, we have not
detected any source with $3 \sigma$ significance.

 To obtain flux, we have produced spectra for the afterglows from
LECS and MECS first TOO data. For absorption and spectral index,
we have selected those with more than 150 photons in the MECS (
background subtracted ). 5 GHOSTs and 9 OTGRBs passed this
criterion.

 We have generally taken the LECS data between 0.1 and 4.0 keV and
the MECS data between 1.6 and 10 keV. The backgrounds we have used are the library ones because they have a very
good signal-to-noise ratio, due to long exposition\footnote{ In the case of GRB970111, 970402 and 991014 the use
of local background enabled us to gather better results.}. However, we have taken the minimum energy for LECS to
be 0.4 keV if the Galactic column density was $N_{H}\ge 5\times 10^{20}$ cm$^{-2}$ because in this case the low
energy backgrounds differ from the library ones, which have been taken at high Galactic latitudes and lower
column densities \citep{Str02}. If we had not adopted this criterion, our analysis would have led to
overestimate the true absorption at the source.

 The standard spectrum model to fit the data consists of a
constant, Galactic absorption, extragalactic absorption (i.e. {\em
in situ}) and a power law. The constant has been included because
LECS and MECS observe a decaying source at different times. Its
value is allowed to vary within a range, obtained in each case by
fitting LECS and MECS data in the 1.6 - 4 keV interval (to avoid
absorption effects) with a simple power law model. The redshift in
our fits has been forced to be 1 for all bursts. This value
corresponds roughly to the average redshift of OTGRBs. We have
adopted this "working hypothesis" to obtain a homogeneous set that
allows us to compare the absorption properties of dark GRBs in the
assumption that they are at the same distance.

 We have calculated the 1.6 - 10 keV flux of dark and bright GRBs
11 hours after the burst trigger. We have chosen this time to
avoid effects of changes in decaying slope. The average count rate
in the MECS has been associated with the flux given by the
spectrum. Successively, we have taken the count rate at 11 hours,
which is given by light curves, to compute the flux at that time.
In most cases, observations include it. In a few cases (e.g.
GRB000926) the flux has been extrapolated.

 For GRB990907, the counts collected were very few and we have
not been able to do any spectral analysis. We have estimated the
flux assuming a spectral index $\alpha = 1.05$. For the two
non-detections, we have calculated the $3\sigma$ upper limits on
counts and converted them to flux adopting again $\alpha = 1.05 $.
In all successive analysis, upper limits have been included as
true afterglows as well as candidate afterglows.

 As a first assessment of our study, we can say that X-ray afterglows
follow the prompt gamma emission in 26 of 31 cases, which
constitute 84\% of the sample. If all doubtful sources are
considered as afterglows, then the fraction of X-ray afterglows
increases up to $94\%$. Instead, optical afterglows are 11 and
constitute only $37\%$\footnote{ GRB980515 has not been included
in this calculation, see further. } of the sample. We note that
all these fractions are in agreement with published data.

 We do not know any optical study on GRB980515. We have calculated
its X-ray flux but this burst has not been included in our
successive analysis.

\section{ The X-ray Spectral and Flux Properties }

 The data we have obtained are the result of the convolution of
the intrinsic distribution with the measurement error distribution. Under the assumption that both are gaussian,
it is possible to deconvolve the two distribution. We have followed a maximum likelihood method \citep{Mac88} to
gather jointly the best estimates of parent distribution mean and standard deviation.
 We have used these best estimates (hereafter, indicated with index {\em m})
for successive analysis, but we have calculated and shown also the
weighted mean and standard deviation of our data. The complete set
of fit parameters is given in Table 1 and plotted in Figures 1 and
2.

  For GHOSTs, the weighted mean and the standard deviation of the
measured energy indexes are $\alpha = 1.3\pm0.17$ (hereafter errors are at $1\sigma$, unless otherwise
indicated) and $\sigma = 0.3$ respectively. The best estimates for the parent population are $\alpha^{m} =
1.3^{+0.27}_{-0.26}$, $\sigma^{m}=0^{+0.4}$ \footnote{In a few cases, the best estimates of the standard
deviation in the parent population are equal to or compatible with zero. This suggests that measurements are
dominated by experimental errors.}. In the case of OTGRBs, $\alpha = 1.04\pm0.03$, $\sigma = 0.44$ and
$\alpha^{m} = 1.05^{+0.11}_{-0.06}$, $\sigma^{m} = 0.05^{+0.13}_{-0.05}$ for the observed and the parent
distribution respectively. Energy indexes of dark and optically bright burst are compatible at $1 \sigma$ level.

 The mean value and the standard (linear) deviation of the measured
absorption (hereafter, in units of $10^{22} {\rm cm}^{-2} $) are, respectively,  $N_{H} = 0.13^{+0.42}
_{-0.13}$, $\sigma = 3$ for dark GRBs, and $N_{H} =0.13\pm 0.06$, $\sigma = 1.7$ for OTGRBs. The best estimates
for the parent population are $ N_{H} ^{m} = 0.14^{+1.46}_{-0.14} $, $ \sigma^{m} = 0^{+1.58} $ for dark GRBs,
and $N_{H}^{m} = 0.13^{+0.13} _{-0.075}$, $\sigma^{m} = 0^{+0.35}$ (see also Stratta et al. 2002) for OTGRBs.
{\em The amount of absorption does not appear statistically different for optically bright and dark GRBs in the
assumption that they lie at the same average $z$.}

 The logarithmic weighted means and the standard deviations of
the observed X-ray fluxes (c.g.s. units) are $<$ log $F$ $>$ $= -12.38\pm{0.02}$ , $\sigma = 0.34$ for dark GRBs
and\\ $<$log $F$ $>$ $= -11.45\pm{0.01}$ , $\sigma = 0.65$ for OTGRBs. Best estimates for the parent population
are $<$log $F$ $>^{m}$ = $-12.53^{+0.11} _{-0.09}$, $\sigma^{m}= 0.23^{+0.09}_{-0.05} $ for dark GRBs and $<$
log $F>^{m}$  = $-11.85^{+0.22} _{-0.23}$, $\sigma^{m} = 0.47 ^{+0.2} _{-0.12}$ for GRBs with OTs. The GHOST
mean flux is likely overestimated, because we have considered upper limits as detections.

  The logarithm of ratio between the mean fluxes of the two parent
populations is $0.68\pm 0.25$, which corresponds to 4.8 in linear
units. A K-S test performed on the flux distributions shows that
the probability that optically bright and dark GRBs derive from
the same population is $P = 2\times 10^{-3}$.  This is a
conservative result, because it has been obtained by including the
upper limits and the non-fading sources as true afterglows in the
set of dark GRBs. If we were to substitute the the non-fading
source fluxes with the $3\sigma$ upper limits of their WFC error
boxes, then the distributions of dark and optically bright GRBs
would be even more different because limits are lower.

\section{Discussion}

 {\em Our analysis shows that dark GRBs have on average weaker X-ray flux
than bright GRBs.} Then, we could simply explain why we miss their
optical detection by assuming that dark bursts are weaker than OT
GRBs in the optical band by the same ratio. Dark bursts should
have had OTs at least 2 magnitudes fainter than OT GRBs; the 4.8
flux ratio that we have found corresponds to $\simeq 1.7$
magnitudes.

 In order to check the viability of this hypothesis, we have
calculated the optical flux density in the R band and hence the optical-to-X flux ratios (hereafter $f_{OX}$) of
each OTGRB and GHOST 11 hours after the burst (Lazzati et al. 2002, Fynbo et al. 2001 and reference therein;
Vreeswijck et al. 1999; Masetti et al. 2001). Upper limits on optical fluxes of GHOSTs have been extrapolated
from the tightest constraint available and adopting an optical flux decaying index $\delta = 1.15 $. Results are
shown in Table 1 and plotted in Figures 3, 4 and 5. We note that the optical and X ray fluxes of OTGRBs {\em
 are correlated:} the higher is the the X-ray flux, the more
luminous is the optical counterpart. The probability that it occurs by chance is only $\thicksim1.5\%$. The
logarithmic standard deviation of $f_{OX}$'s is $\sigma_{f_{OX}} = 0.42^{+0.2} _{-0.12}$, which corresponds to a
multiplicative factor of 2.6, while the logarithmic mean is $<$ log $f_{OX} ^{ OT}$  $>^{m}$ = $0.3\pm 0.22 $ if
the X-ray and optical fluxes are expressed in $10^{-13}$ erg cm$^{-2}$ sec$^{-1}$ and $\mu$Jy respectively.  We
have fitted the distribution of X-ray and optical fluxes with the function log $F_{optical}$ = K + $A$
log(F$_{1.6 - 10 keV)}$. The best fit values are $A = 0.81 $, K = $0.41$.
 We have also calculated the average optical-to-x spectral
index, $\overline{\alpha}_{OX}$, as function of $<$ log $\overline{f}_{OX}^{ OT}$ $>^{m}$, by adopting the X-ray
and optical density flux at 2 keV and R band respectively and X-ray spectral index $\alpha_{X} = 1.05$. Our
result is $\overline{\alpha}_{OX} ^{ OT} = 0.79\pm 0.054$.

  If we exclude GRB980519, which seems to be the only afterglow
explained by interaction of a jet outflow with a star wind medium \citep{Jau00}, the correlation is
strengthened: $\sigma_{f_{OX}} = 0.28^{+0.14} _{-0.08} $, which corresponds to a multiplicative factor of 1.9;
the probability of a chance occurrence is $ < 0.001 $; $<$ log $ f_{OX} ^{ OT}>^{m}$ = $0.18^{+0.16} _{-0.14} $.
The best fit values are $A = 0.91 $, $K = 0.24$.

 We can immediately recognize that $75\%$ of the GHOSTs of
our sample (14 of 19) have optical flux upper limits consistent with OT detections (see Fig. 3), so that they
may not be actually "dark". Optical follow-ups conducted for these bursts would not have been deep enough to
detect the faintest OTs in our set. A similar fraction has been found out by \citet{Fyn01a} and \citet{Ber02},
comparing sets of non-detections with the light curves of the dim afterglows of GRB000630 and GRB020124. It is
worth noting that the $f_{OX}$ upper limits of these 14 objects are quite similar to OTGRBs ones: a K-S test
performed shows that the probability they belong to the same population is not marginal\footnote{GRB010220 has
limits on optical and X-ray flux, so its $f_{OX}$ is not constrained. However, wherever it is, it would not
affect much the result.}. This result gives support to the fact that they could be faint sources with optical
properties assimilable to X-ray ones.

 The remaining 5 objects, which constitute $25\%$ of GHOSTs,
have optical flux upper limits lower than all OTGRBs in our set (see Fig. 2). Furthermore, their optical
emission must have been even 2-3 times fainter than the dim afterglow of GRB000630 and 4-6 fainter than
GRB020124 \citep{Ber02}, which had an R band flux $F_{11h} ^{R} \simeq 3.4 \mu$ Jy  and $F_{11h} ^{R} \simeq 7.9
\mu$ Jy respectively \footnote{ Data extrapolated with best fit values given by the authors and corrected for
Galactic extinction.}. These two optical afterglows, however, are not the weakest ever occurred. In our set, the
OT of GRB980613 is even dimmer (see table 1) and establishes a more stringent test.

 We wonder if the hypothesis of weaker flux at all wavelenghts
can hold for these 5 GHOSTs (hereafter, we will refer to them as
"the darkest" or the "most obscure", etc. for simplicity). If so,
we would expect that their X-ray fluxes were proportionally very
weak like the optical fluxes, so that their $f_{OX}$'s should be
not very different from OTGRBs.
 In 4 cases of 5 the $f_{OX}$'s are lower than all OTGRBs and $ 4 - 10 $
times lower than the average optical-to-x flux ratio of OTGRBs. The exception is GRB990217, which has upper
limits both in optical and in X-rays flux. If we use both of them, then we get log $f_{OX} = 0.2 $ which is much
more similar to $<$ log $f_{OX} ^{ OT}$ $>^{m}$. We have performed a K-S test on $f_{OX}$ between all these
darkest bursts and all OTGRBs. The probability that they are drawn from the same distribution is $ P \leq 0.01 $
($\geq2.6\sigma$ confidence level). The average optical-to-X spectral index of these objects
$\overline{\alpha}_{OX} \leq 0.62$, well below that of OTGRBs. Therefore, we have a strong indication that for
these bursts the spectrum is depleted in the optical band, by $\sim 2 $ magnitude on average.

 The x-ray mean flux of these 5 GHOSTs is $-12.48\pm{0.16}$. The logarithmic
ratio between the OTGRB X-ray flux mean and this mean is log $ r =
0.63 \pm 0.28$, which corresponds to a 4.3 factor. A hypothesis
for the absence of OT and fainter X-ray flux is that of very high
redshift. The GRB prompt emission and X-ray afterglow of the
strongest bursts (e.g. GRB990123 and GRB990510) could be
detectable even they occur at $ z > 10 $ \citep{Lmb00a}. However,
if GHOSTs were at $ z \gtrsim 5 $, then extragalactic hydrogen
clouds would entirely wash out optical emission
\citep{Pir02b,Fru99,Bec01}.

 To estimate the average redshift of the most obscure GRBs, we shall
use the formula \citep{Lmb00a}:

\begin{equation}
 F(\nu,t) = \frac{L_{\nu}(\nu,t)} {4\pi D^2(z) (1+z)^{1-\alpha+\delta}}
\end{equation}

 where $\alpha$ is the spectral index, $\delta$ is the decaying (temporal)
index and $D(z)$ is the comoving distance. We assume the
cosmological parameters values $H_{ 0} = 65 {\rm km}\; {\rm
sec}^{-1} \;{\rm Mpc}^{-1} $, $\Omega_{M} = 0.3$,
$\Omega_{\Lambda} = 0.7$. The average of the known redshifts of
OTGRBs in our set is $\overline z_{OT} \backsimeq 1.5 $.

 In the simplest model of GRB afterglows,
$ \delta = -4/3 , \alpha = 2\delta /3$, so $ 1 - \alpha + \delta =
5/9$. For such parameters, the average redshift of the darkest
GRBs should be $2.6 \leq \overline{z} _{D} \leq 8.7$ under the
assumption that the lower mean flux were only due to their larger
distances and not to an intrinsic difference in their luminosity.
Using the best estimate of $ \alpha = 1.05 $ calculated for OTGRBs
and the average $\delta = 1.33 $ of the strongest bursts of our
sample, we obtain $2.3\leq \overline{z} _{D} \leq 7.8$. We should
also expect a distribution of burst redshifts around $\overline{z}
_{D} $. These facts make the high redshift scenario for the most
obscure GHOSTs still plausible.

 Adopting the hypothesis GRBs are the final result of very massive
star evolution, an interesting issue to address is what might be the
progenitors of GRBs at very high redshifts.

 Currently, we observe only old and low-mass Population II stars, but even
high mass stars could have formed. Theories suggest that the first stars of
the Universe - the so-called Population III - might have very large mass, so
they could possibly be good candidates.
 Recent calculations suggest \citep{Lmb00b,Val99,Gne97,Ost96} that the star
formation rate has two peaks. The first one, at $ 20\gtrsim z
\gtrsim 16 $ is due to Population III stars. The second one, due
to Population II, is higher and much broader and it is at a
redshift in the range $ 12 \gtrsim z \gtrsim 2 $. Also the number
of stars (i.e. the star formation rate time-dilated and weighted
by the comoving volume of the universe) shows two peaks at $ z
\thicksim 8 $ and $ z \thicksim 2 $.

 In a few cases, however, the redshifts of some dark GRBs have been almost
securely found, e.g.  GRB970828 at $z=1$ \citep{Djo02} , GRB
000210 at $z=0.85$ \citep{Pir02}, GRB 000214 at $z=0.44$
\citep{Ant00}, while GRB981226 is also likely to have not occurred
at very high redshift \citep{Fra99}, because the candidate host
galaxy is still detected in the R band. With present statistics,
at least $\thicksim 15\%$ of the examined dark GRBs are not at
very large redshift. It should be noticed that two of them are
included in the list of most obscure objects in our set.

 An hypothesis to explain the lack of the optical emission, alternative
to the very high redshift scenario, may be strong absorption \citep{Djo01}. So far, we have collected many
indications that GRBs take place in dense environments, like the Giant Molecular Clouds (hereafter GMCs)
\citep{Pir02b}. GMCs are very rich in dust, which extinguishes very efficiently the optical and UV light.
\citet{Pir02} argues that in the case of GRB000210 the lower limit on amount of obscuration is 1.6 magnitude in
the R band. This value has been obtained extrapolating a power law spectrum, described by the fireball model,
from the X-ray band to the optical band and comparing the expected flux with the upper limits. We find a similar
result through our model-independent analysis of optical-to-x flux ratios. The $f_{OX}$ upper limits of the
burst is 3.8 times lower than $\overline{f}_{OX} ^{ OT}$, which corresponds to $\gtrsim 1.5$ magnitude
depletion. The measurements of Chandra X-ray Observatory showed that the amount of local absorbtion is able to
explain this obscuration, under the assumption that the dust-to-gas ratio of the intervening medium is the same
of the Galaxy. However, we note that in the case of OTGRBs the dust-to-gas ratio seems not to be consistent with
the Galactic one \citep{Str02,Gal01}. Similarly, \citet{Djo02} derived extinction in the case of GHOST GRB970828
\citep{Yos01}, for which a significative amount of X-ray absorption was detected.

 If the most obscure GHOSTs were similar to GRBs with OT except
for higher absorption, we would expect to see differences in values of $N_{H}$. From our results, we cannot
affirm that $N_{H}$ in these bursts shows this tendency, also due to considerable errors (see Table 1 and Figure
1). For those with good statistics, we do not find any absorption value $ 3 \sigma $ higher than the Galactic
value but marginal evidence ($\sim 2 \sigma$). On the other hand, we cannot rule out the hypothesis of obscuring
GMCs altogether. The upper limits on $N_{H}$, a few $ \times 10^{22}$, are in fact the typical column densities
of GMCs. {\em The optical absorption, however, does not imply that most obscure GHOSTs have X-ray flux weaker
than OTGRBs, as we have found in our analysis,} because X-ray absorption is almost negligible at energy larger
than 1.6 keV. Reichart \& Yost (2001) try to reconcile this fact with the hypothesis of dusty birthplaces for
GRBs and, in particular, they considered the effect of variously beamed GRB fireballs on their dusty
environments. The energetics of GRBs are more or less the same for all events \citep{Fra01}, but the beaming
angles differ, being narrower for stronger bursts. The larger the beaming angle is, the more difficult it is for
the diluited prompt UV and X-ray emission to destroy dust along the line of sight (Waxman \& Draine 2000;
Fruchter et al. 2001; Draine \& Hao 2002; Perna \& Lazzati 2002), so that we see a weak GRBs without OT. With a
narrower beaming angle, the prompt emission will destroy a larger fraction of dust and the GRBs will appear
strong and with OT. If this hypothesis is correct, on the basis of our results we have to assume that the
average beaming angle of the darkest GHOSTs is $\sim 2$ times wider than the OTGRB one. According to Frail et
al., the average beaming angle of BeppoSAX OT GRBs is $\theta \sim 0.1\; {\rm rad}$, so that the average beaming
angle of the darkest GRBs should be $\theta \sim 0.2\;{\rm rad} $. This prediction is important, because it can
be experimentally tested by observing and timing the presence of achromatic breaks in the light curves.

 Another consequence of dark GRB occurrence in high density environments
should be the detection of semi-ionized absorber in the low energy
X-ray spectrum. So far, this kind of feature has not been found.
The ionization front, however, should be rather sharp (see e.g.
Draine \& Hao 2002), and therefore it would be hard to detect
signatures of semi-ionized species in the X-ray spectra of the
bursts.

\section{Conclusions}

 We have discussed the issue of GRBs with X-ray but no optical afterglows.
 We have performed a standard temporal and spectral analysis of a complete sample
of 31 GRB X-ray follow up observations of BeppoSAX, i.e. all the
fast observations from the launch until February 2001. We have
found that X-ray afterglows follow the prompt gamma emission in
$84\% - 94\% $ of the cases.

 We have obtained the 1.6 - 10 keV fluxes 11 hours after the
trigger for each GRB and the values of N$_{H}$ at $z=1$ to compare
the absorption properties for strong X-ray afterglows. While
absorption of optically bright and dark GRBs does not appear to be
significatively different, the fluxes of GRBs with OT are on
average about 5 times stronger than GHOST ones. The probability
that GHOSTs and optically bright GRBs belong to the same
population in fluxes is $\leqslant 0.002$.

 From the very fact that X-ray flux of dark GRBs is 5 times lower than OTGRBs,
the optical flux could be $\sim 2$ magnitude lower under the assumption that the shape of the optical-to-X
spectrum is the same of OTGRBs. This difference could explain the non-detection of the optical transient. In
order to test this hypothesis, we have calculated the optical-to-X flux ratios of OTGRBs and upper limits for
GHOSTs. OTGRBs show a tight correlation of optical and X-ray fluxes. The mean for OTGRBs is $<$ log $f_{OX} ^{
OT}$ $>^{m}$ = $0.3\pm 0.22$ and $\sigma = 0.4 $ ; the probability of a chance correlation is a marginal $\sim
1\%$. We find that $75\%$ of GHOSTs have $f_{OX}$ upper limits similar to OTGRB ones; however, the remaining
$\sim 25\%$ of dark bursts are fainter in optical than in X-rays, being their average optical-to-x flux ratio
$<$ log $f_{OX} $ $>^{m}$ $\leq -0.4$. Thus, we have a strong indication that for these bursts the spectra are
different from OTGRBs. This result is significant at $\geq 2.6 \sigma$ level.

 Two different interpretations for this effect can be given:
1) location at $z>5$ ; 2) higher absorption. In the very high
redshift scenario, the optical flux of the sample is extinguished
by the intervening Ly-$\alpha$ systems, while the X-ray flux lower
than OTGRBs is understood in terms of a higher distance.

 However, given the fact that some GHOSTs of the sample almost
certainly do not lie at very high redshift, we have considered the alternative possibility of occurrence in
dusty and dense environments like GMCs. We have not found that these bursts have a higher absorption than
optically bright GRBs, but we note that upper limits on $N_{H}$ are consistent with those of giant clouds. In
the case of GRB000210 our model-independent analysis has shown a depletion in the optical which is compatible
with the X-ray absorption measured by Chandra, assuming a dust-to-gas ratio similar to that of our Galaxy.

 In the near future, a key role will be played by fast and deep
follow up X-ray and optical observations of GRBs, which will allow
us to constrain better their spectral properties. In particular,
observations in the IR band are a really important tool because
they are less sensitive to dust and to Ly-$\alpha$ extinction.
They will enable us to investigate dark GRB properties like
distance, that is a crucial piece of information to disclose the
nature of these objects.

\acknowledgements

 M.D.P. is grateful for support from University "La Sapienza" grant.
BeppoSAX was a joint program of Italian Space Agency (ASI) and
Netherlands Agency for Aerospace Programs (NIVR). We warmly thank
all members of BeppoSAX Scientific Operation Center and Operation
Control Center.

{}
\textbf{}

\clearpage
\begin{table}
\begin{center}
\caption{GRB x-ray flux and optical density flux, spectral index $\alpha$, absorption at $z = 1$, optical-to-X
flux ratio. Errors at 90\% Confidence Level.\label{tbl-2}}
\begin{tabular}{rccccc}
\tableline\tableline GRB & X-ray Flux \tablenotemark{a} & $\alpha$
&$N_{H} (10^{22}$ cm$^{-2}$)   & \multicolumn{1}{c}{
$f_{OX}$\tablenotemark{b} } &
Optical Flux ($\mu$Jy) \\

\tableline

& & & & & \\
\vspace{-0.55cm}
Dark GRBs& & & & & \\
\vspace{0.1cm}
& & & & & \\
970111 &$1.11\pm 0.35 $\tablenotemark{c} & & & $\leq 27.4$ &$\leq 30.4$\\

970402 & $2.62 \pm 1.31            $ &    &  &$\leq 7.82$ &$\leq 20.5$\\

971227 &$3.24   ^{+1.59}  _{-2.08} $ &    & &$\leq 1.5$ &$\leq 4.87$\\

980515 &$2.01   ^{+0.54} _{-0.93}  $ &    & & &\\

981226 &$4.88  ^{+0.4}   _{-0.73}  $ &    & &$\leq 0.32$ &$\leq 1.56$\\

990217 &$\leq1.11 \tablenotemark{d}$ &    & &$\leq 1.6 $&$\leq 1.77$\\

990627 &$1.87  ^{+0.83}  _{-1.08}  $ &    & &$\leq 16.9$ &$\leq 31.6$\\

990704 &$5.95  ^{+1.29}  _{-1.29}  $ & $1.75 ^{+1.09} _{-0.59} $ &
$ 4.83^{+10.37}_{-3.57} $ &$\leq 0.2$ &$\leq 1.19$\\

990806 &$3.8  \pm 1.03   $ &     $ 1.56 ^{+1.03} _{-0.71} $ &
$3.16^{+10.64} _{-3.09} $ &$\leq 0.4$ & $\leq 1.5$\\

990907 &$10.2    \pm 5.6           $ &    & &$\leq 0.78$ &$\leq 8$\\

991014 &$4.01  ^{+1.37} _{-1.2}    $ &    & &$\leq 0.89$ &$\leq 3.6$\\

991106 &$2.09  \pm 1.08 \tablenotemark{c} $ &    &  &$\leq 12.6$ &$\leq 26.3$\\

000210 &$3.69  ^{+1.02}  _{-1.08}  $ & $ 1.67 ^{+1.01} _{-0.78} $ &
$2.95^{+6.3} _{-2.27}$ &$\leq 0.52$ &$\leq 1.92$\\

000214 &$6.37  ^{+1.98} _{-1.77} $   & $ 1.18 \pm{0.43} $ & $0
^{+0.71}$ &$\leq 7.59$ &$\leq 48.4$\\

000528 &$2.33   \pm   1.04       $  &    &    &$\leq 1.31$ &$\leq 3.05$\\

000529 &$3.55  ^{+1.24} _{-2.16} $  &    &    &$\leq 11.91$ &$\leq 42.3$\\

000615 &$1.28\pm 0.33 \tablenotemark{c}$ & & &$\leq 2.04$  &$\leq 2.61$ \\

001109 &$20  ^{+5.8}  _{-4.6}    $ & $ 1.26 ^{+0.12} _{-0.49} $
&$2.83^{+4.7}_{-2.83} $ &$\leq 0.59$ &$\leq 11.81$\\

010214 &$2.67 ^{+0.93}  _{-1.25} $ &     &   &$\leq 1.89$ &$\leq 5$ \\

010220 &$ <1.63 \tablenotemark{d}$ & &  &$\leq 14.5$  & $\leq 23.2$ \\

& & & & & \\
\vspace{-0.55cm}
OT GRBs & & & & & \\
\vspace{0.1cm}
& & & & & \\

970228 &$19.7\pm{3.3}          $ & $0.8  ^{+0.3} _{-0.37}$
&$0.83^{+.1.51}_{-0.83} $ &$2.2$ &$43.8^{+5.5} _{-4.9}$\\

970508 &$7.91\pm{0.67}         $ & $1.14  ^{+0.51} _{-0.36}$
&$0.53^{+1.87} _{-0.53} $  &$1.26$ &$9.6^{+0.74} _{-0.71}$\\

971214 &$6.03\pm{1.09}           $ & $0.98 ^{+0.44} _{-0.56}$
&$2.98^{+6.51} _{-2.98}$   &$0.86$ &$5.2\pm{0.56}$\\

980329 &$5.99\pm{0.93}   $ & $1.42 ^{+0.58} _{-0.39}$
&$0.21^{+4.05}_{-0.21}$  &$0.67$ &$4^{+2.4} _{-1.3}$\\

980519 &$3.97\pm{0.92}           $ & $2.2  ^{-1.55} _{+1.09}$
&$3.2^{+11.5} _{-3.2} $ &$20.6$ &$82 ^{+10.4} _{-9.2}$\\

980613 &$2.14\pm{0.86}           $ &   &   &$1.14$ &$2.4^{+2.4} _{-1.2}$\\

980703 &$15.6 ^{+7.7}   _{-5.6}   $ & $1.77 ^{+0.6} _{-0.47} $
&$2.88^{+4.74} _{-2.06} $   &$4.34$ &$67.7\pm{28.8}$\\

990123 &$53 \pm 2           $ & $0.99  ^{+0.07} _{-0.08} $     &$
0.09^{+0.11} _{-0.05} $  &$0.92$ &$40.33\pm{0.93}$\\

990510 &$36.7 \pm{2.8}            $ & $1.19  \pm{0.14} $        &$
0.21^{+0.61} _{-0.21} $ &$4.44$ &$163\pm{15.6}$\\

000926 &$39.6 ^{+22.4} _{-19.1}   $ &    &   &$3.94$ &$156.9\pm{9}$\\

010222 &$68 \pm 4.2               $ & $1 \pm{0.1} $& $0.53
^{+0.42} _{-0.27} $ &$0.74$ &$50.6\pm{2.3}$\\

\end{tabular}

\tablenotetext{a}{ $10^{-13}$ erg cm$^{-2}$ sec$^{-1}$ }

\tablenotetext{b}{ Obtained dividing the R band flux ( or upper
limits ) in $\mu$ Jy by the $1.6 - 10$ keV X-ray flux in $10^{-13}
$cgs .}

\tablenotetext{c}{ Candidate afterglow }

\tablenotetext{d}{ $3 \sigma$ upper limit }

\end{center}
\end{table}

\clearpage

\begin{figure}
\includegraphics[width=15.3cm,height=15.3cm]{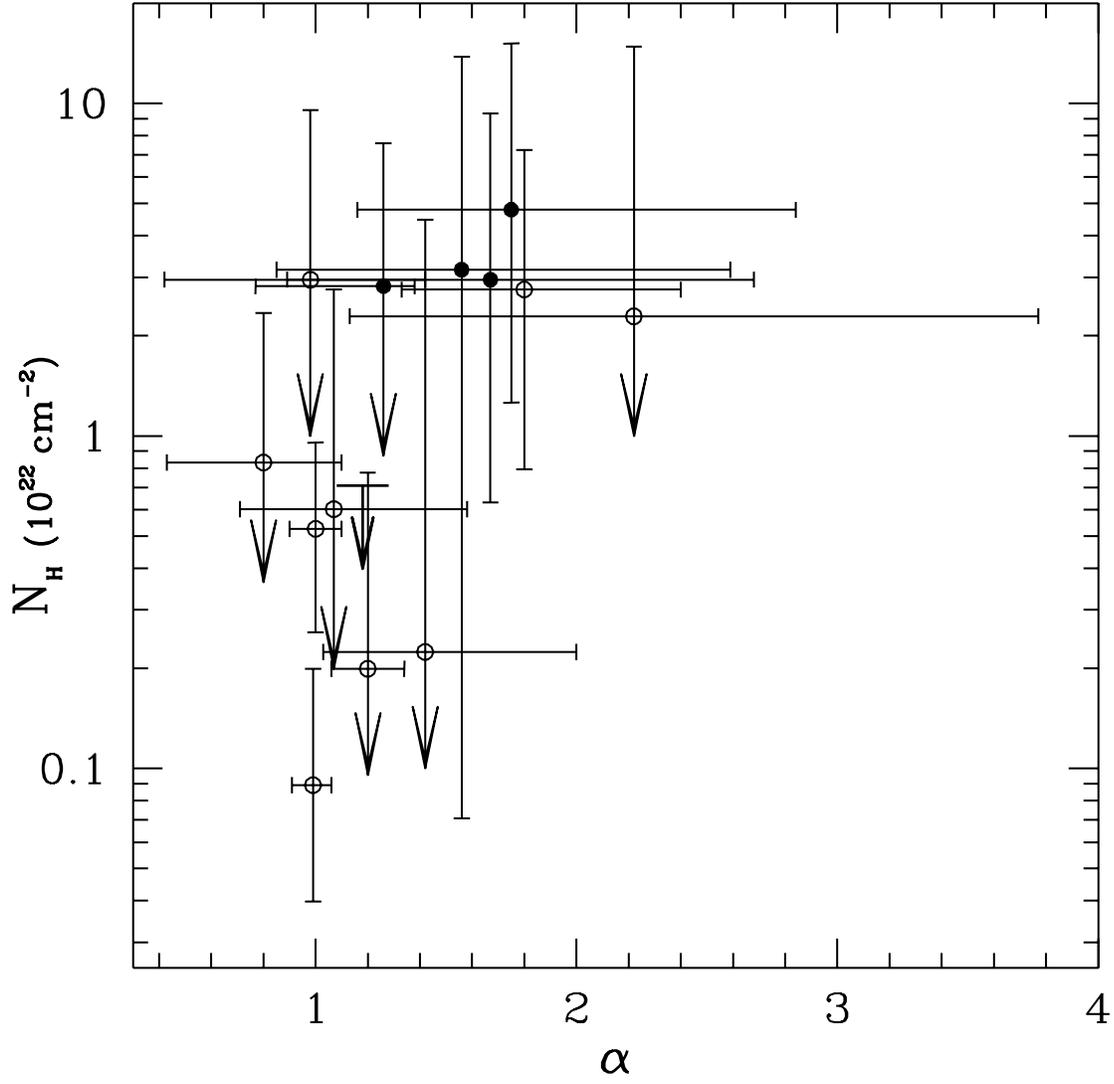}
\caption{ $N_{H}$  vs spectral index of high statistics GRBs. Filled dots: dark GRBs. Empty dots: OTGRBs. }
\label{fig1}
\end{figure}

\begin{figure}
\includegraphics[width=15.3cm,height=15.3cm]{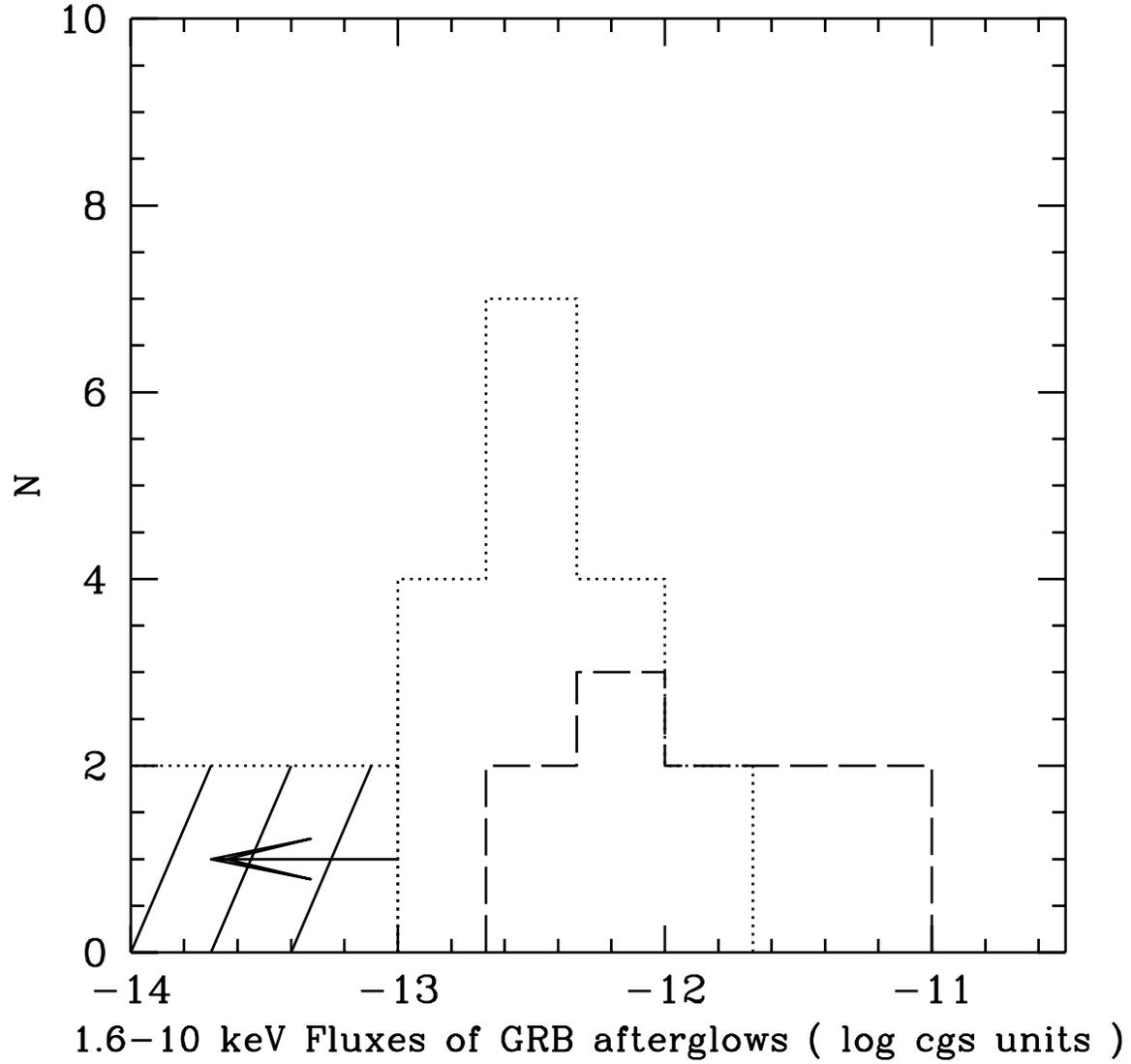} \caption { Hystogram of 1.6 - 10 keV fluxes of GRBs
11 hours after the burst. Long-dashed line:
OT GRBs. Dotted line: dark GRBs, candidate afterglows included. The arrow indicates the two upper limits set
$\equiv 10^{-13}$ in order to clarify the picture. } \label{fig2}
\end{figure}

\begin{figure}
\includegraphics[width=15.3cm,height=15.3cm]{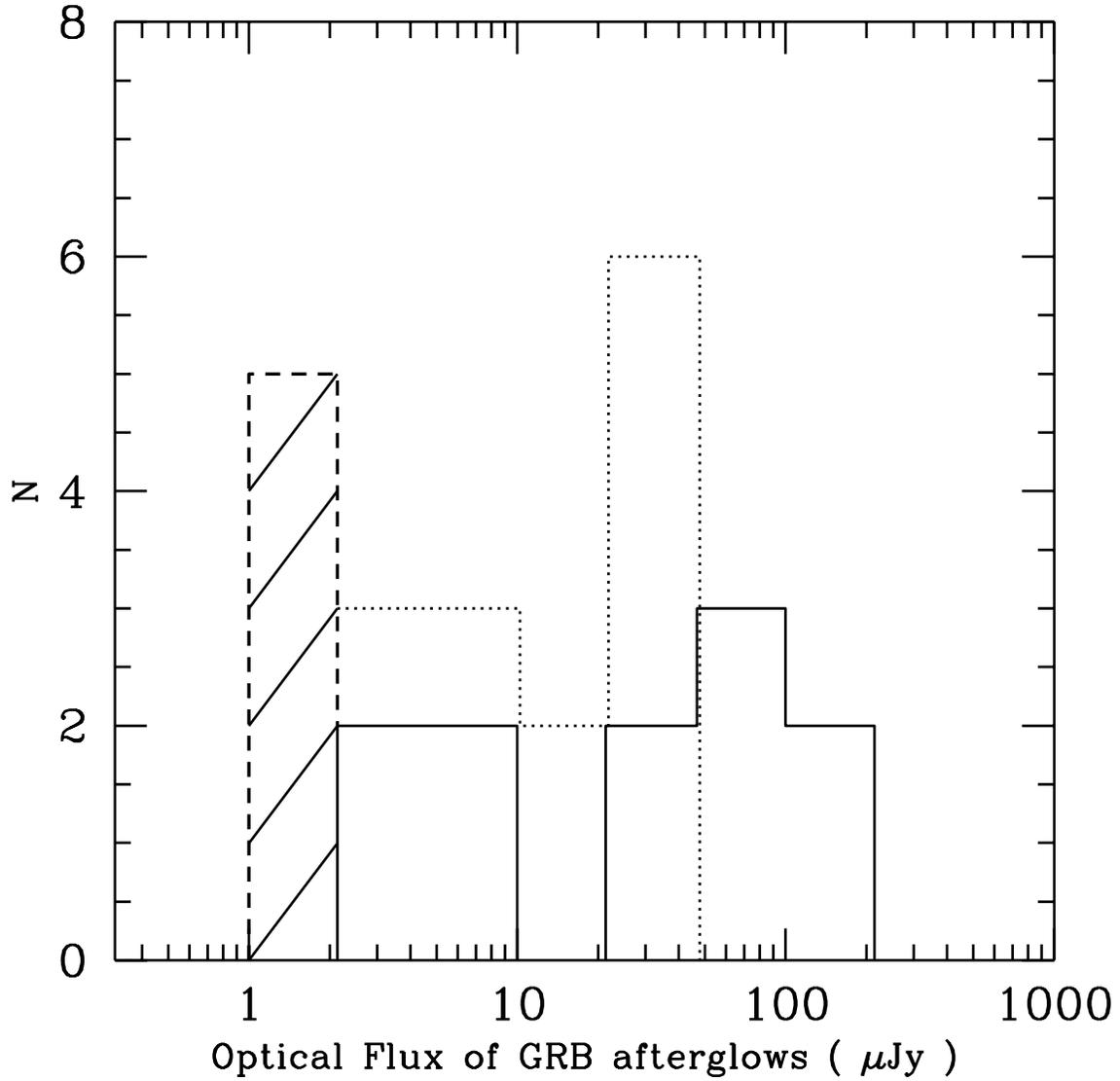} \caption { Hystogram of all the GRB optical fluxes and upper limits 11 hours after the burst.
Solid line: OTGRBs. Dotted line: GHOST upper limits. Short-dashed line: the most obscure GHOSTs.} \label{fig3}
\end{figure}

\begin{figure}
\includegraphics[width=15.3cm,height=15.3cm]{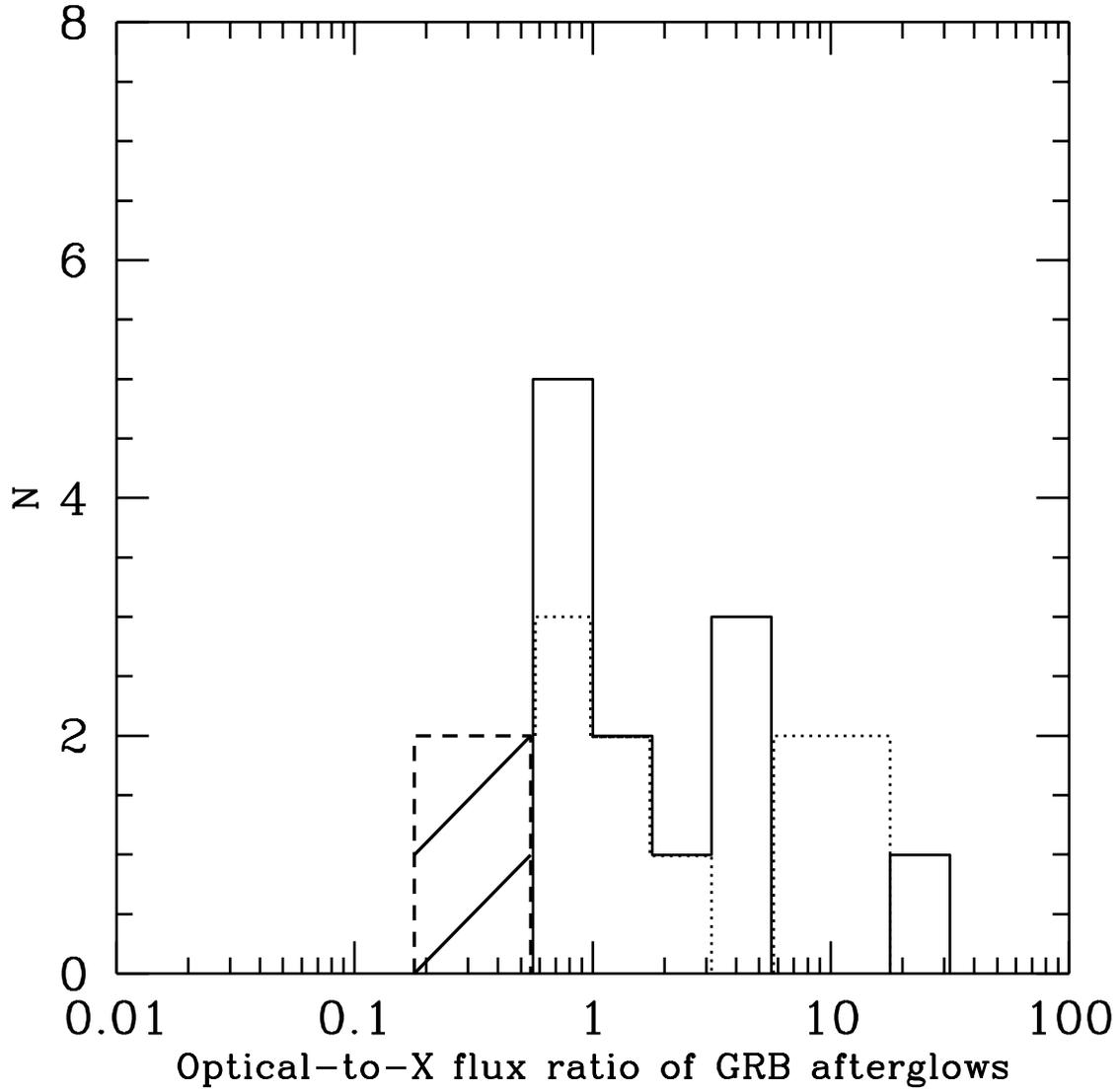} \caption { Optical-to-X flux ratios. Dotted lines: dark GRB upper limits. Short dashed: the most
obscure dark GRBs. Solid line: OTGRBs. Non-detected X-ray afterglows are not shown. } \label{fig4}
\end{figure}

\begin{figure}
\includegraphics[width=15.3cm,height=15.3cm]{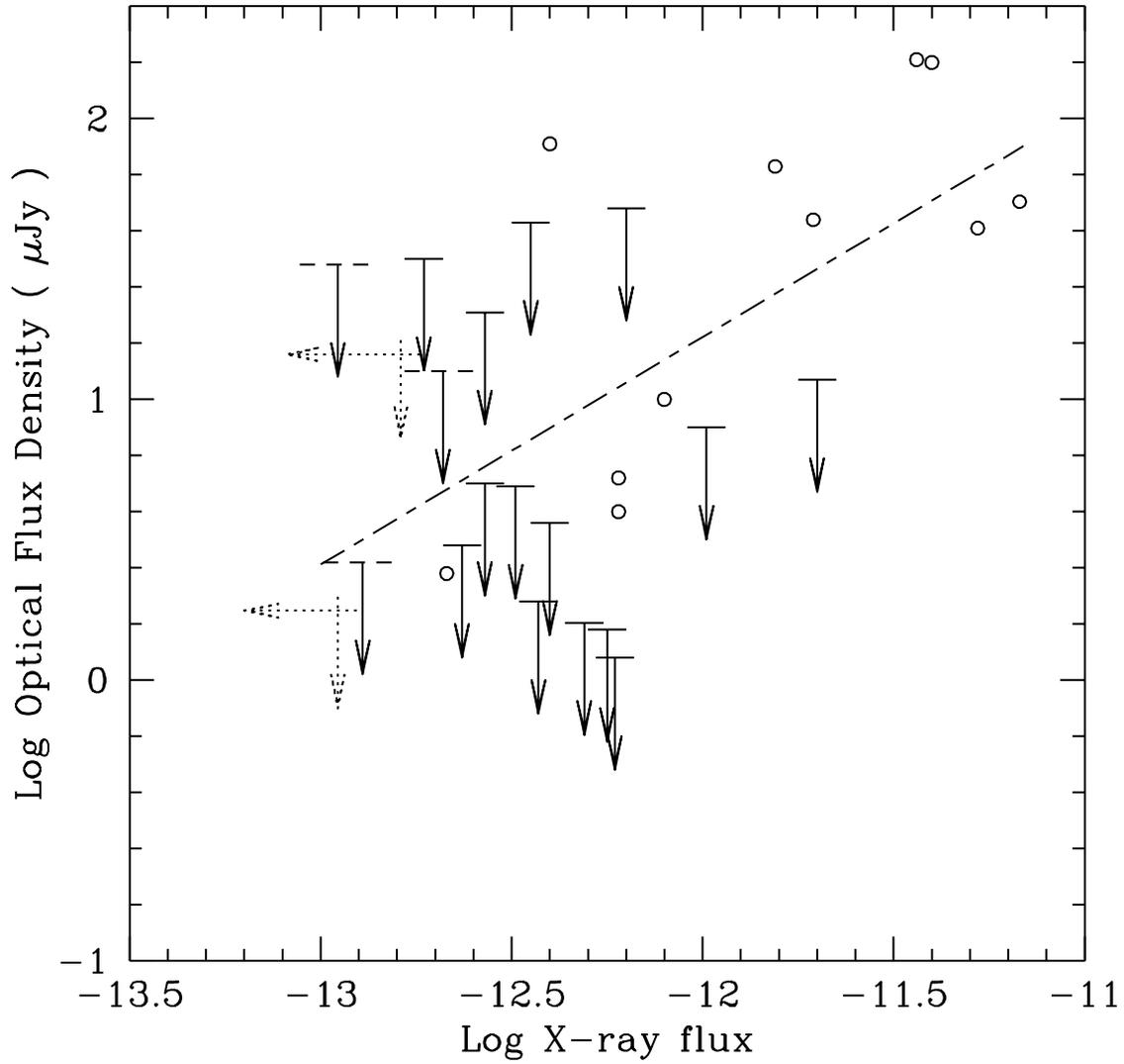} \caption { X-ray vs optical flux of GHOSTs and OTGRBs. Empty dot: OTGRBs. Solid arrows: GHOSTs.
Dashed arrows: candidate sources. Dotted arrows: upper limits. Short-long dashed line: best fit of optical vs X
flux for OTGRBs, GRB980519 included.} \label{fig5}
\end{figure}

\end{document}